\documentclass[a4paper]{jpconf}
\usepackage{graphicx}
\begin{document}
\title{Theory of two-dimensional macroscopic quantum tunneling in a Josephson junction coupled with an LC circuit}

\author{Shiro Kawabata$^{1,2,3}$, Takeo Kato$^{4}$, Thilo Bauch$^{2}$}

\address{
$^1$Nanotechnology Research Institute (NRI), National Institute of Advanced Industrial Science and Technology (AIST), Tsukuba, Ibaraki, 305-8568, Japan
\\
$^2$Department of Microelectronics and Nanoscience (MC2), Chalmers University of Technology, S-41296 G\"oteborg, Sweden
\\
$^3$CREST, Japan Science and Technology Corporation (JST), Kawaguchi, Saitama 332-0012, Japan 
\\
$^4$The Institute for Solid State Physics (ISSP), University of Tokyo, Kashiwa, Chiba, 277-8581, Japan
}

\ead{s-kawabata@aist.go.jp}

\begin{abstract}
We investigate  classical thermal activation (TA) and macroscopic quantum tunneling (MQT) for a Josephson junction coupled with an LC circuit theoretically.
The TA and MQT escape rate are calculated analytically by taking into account the two-dimensional nature of the classical and quantum phase dynamics. 
We find  that the MQT escape rate is largely suppressed by the coupling to the LC circuit.
On the other hand, this coupling gives rise to slight reduction of the TA escape rate.
These results are relevant for the interpretation of a recent experiment on the MQT and TA phenomena in grain boundary YBCO Josephson junctions.
\end{abstract}

\section{Introduction}

Recent experimental observations of macroscopic quantum tunneling (MQT)~\cite{rf:Bauch1,rf:Inomata,rf:Matsumoto,rf:Li,rf:Kashiwaya} and energy level quantization~\cite{rf:Bauch2,rf:Jin} in high-$T_c$ superconductor Josephson junctions (JJs) open up the possibility for realizing high-$T_c$ quantum bits.

In YBCO grain boundary junctions which was used in MQT experiments~\cite{rf:Bauch1,rf:Bauch2}, it was found that the stray capacitance $C_S$ of the electrodes, due to the large dielectric constant of the STO substrate at low temperature ($\epsilon_r>10000$), and the inductance $L_S$, due to the large London penetration depth in $c$-axis and/or Josephson coupling between CuO${}_2$ planes in one of the electrodes have large influence on the macroscopic dynamics and can be taken into account by an extended circuit model (see Fig.1(a)).
In Fig. 1, $\phi$ is the phase difference across the Josephson junction and $\phi_s= (2 \pi/\Phi_0) I_S L_S + \phi$ is the phase difference across the capacitor $C_S$, where $I_S$ is the current through the inductor.
It was found that, in the microwave-assisted MQT experiment~\cite{rf:Bauch2}, the bias current $I_\mathrm{ext}$ dependence of the Josephson plasma frequency $\omega_p$ is quantitatively explained by this model~\cite{rf:Lombardi,rf:Rotoli}.
However, the validity of the extended circuit model for TA and MQT escape processes have not yet been explored.
Below we calculate the MQT and TA escape rate base on the extended circuit model and try to compare with experimental results.

\begin{figure}[tb]
\begin{center}
\includegraphics[width=12.0cm]{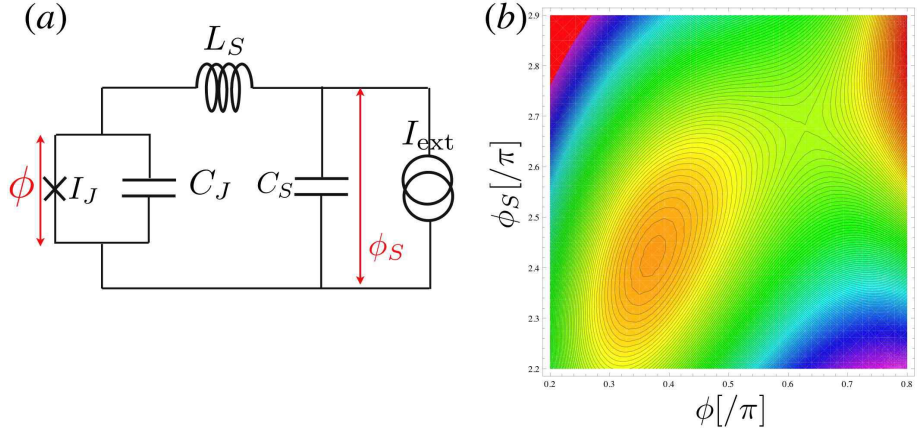}
\caption{(a) Extended circuit model and (b) the contour plot of the potential $U(\phi,\phi_S)$ for $\gamma =0.92$ and $\eta=7$.}
\end{center}
\end{figure}

\section{Model}

The Hamiltonian of the circuit (Fig. 1(a)) can be written as
\begin{eqnarray}
{\cal H}
&=&
\frac{Q_J^2}{2 C_J}
+
\frac{Q_S^2}{2 C_S}
+
U(\phi, \phi_s)
\\
U(\phi, \phi_s)
&=&
E_J 
\left\{ 
-\cos \phi + \frac{\left( \phi - \phi_S \right)^2}{2 \eta} - \gamma \phi_S
\right\}
 ,
\end{eqnarray}
where  $Q_{J}=C_{J}(\Phi_0/2 \pi) (d \phi /d t)$, $Q_{S}=C_{S}(\Phi_0/2 \pi) (d \phi_S /d t)$, $E_J$ is the Josephson coupling energy,  $\eta \equiv 2 \pi I_C L_S / \Phi_0= L_S/L_{J0} (L_{J0} = \Phi_0/2 \pi I_C)$, and $\gamma= I_\mathrm{ext}/I_C (I_\mathrm{ext}$ is the external current).
This Hamiltonian describes the quantum dynamics of a fictive particle moving in the 2-dimensional tilted washboard potential $U(\phi,\phi_S)$ (see Fig.1(b)).
By introducing the new coordinate $(x,y)=(\phi-\phi^m,\phi_S-\phi_S^m)$ and assuming $\gamma \approx 1$, we can rewrite the Hamiltonian as 
\begin{eqnarray}
{\cal H}
&=&
\frac{1}{2}M \dot{x}^2  + \frac{1}{2}m \dot{y}^2  + U(x,y)
\\
U(x, y)
&=&
E_J 
 \left(-\frac{x^3}{6}
+
\frac{\sqrt{1- \gamma}}{\sqrt{2}}  x^2
+
\frac{x^2}{2 \eta}
+
\frac{y^2}{2 \eta}
-
\frac{x y}{ \eta}
\right)
 ,
\end{eqnarray}
where $(\phi^m, \phi_S^m)$ is the local minimum point, $M=C_J (\Phi_0/2 \pi)^2$, and $m=C_S (\Phi_0/2 \pi)^2$.

By using the functional integral method~\cite{rf:Weiss,rf:Kawabata1,rf:Kawabata2,rf:Kawabata3,rf:Kawabata4,rf:Yokoyama,rf:Kawabata5}, the partition function of the system can be written as
$
{\cal Z} 
=
 \int {\cal D} x(\tau)
 \int {\cal D} y(\tau)
\exp
\left[
- \int_0^{\hbar \beta} {\cal L} [x,y] / \hbar
\right]
,
$
where $ {\cal L} [x,y]$ is the Euclidean Lagrangian.
The Lagrangian is a quadratic function of $y$.
Therefore the functional integral over variable $y$ can be performed explicitly.
Then the partition function is reduced to a single functional integral over $x$, $i. e.$, ${\cal Z} = \int {\cal D} x(\tau) \exp (- {\cal S}_\mathrm{eff} [x] /\hbar)$, where the effective action is given by 
\begin{eqnarray}
{\cal S}_\mathrm{eff}[x]
&=&
\int_0^{\hbar \beta}
d \tau \left[
\frac{1}{2} M \dot{x}^2 + U_\mathrm{1D}  (x)
\right] 
+
\frac{1}{4}
\int_{0}^{\hbar \beta} d \tau
\int_{0}^{\hbar \beta} d \tau'
\left[ x(\tau) - x(\tau') \right]^2  K (\tau-\tau') 
.
\end{eqnarray}
In this equation, $U_\mathrm{1D}=E_J \left[  - x^3 / 6 + \sqrt{(1- \gamma)/2} x^2\right]$, and
\begin{eqnarray}
K(\tau)
&=&
\frac{1}{2} m \omega_{LC}^3
\frac{
      \cosh \left[  \omega_{LC} \left(  \frac{\hbar \beta}{2} - \left| \tau \right| \right)\right]
      }
      {
      \sinh \left[  \frac{ \hbar \beta \omega_{LC} }{2}  \right]
      }
      ,
\end{eqnarray}
where $\omega_{LC} = 1/\sqrt{L_S C_S}$ is the resonance frequency of the LC circuit.
Thus the dynamics of the phase difference in the 2D potential $U(\phi,\phi_S)$ can be mapped into simple 1D model.
Note that, due to the coupling between the Josephson junction and the LC circuit, ${\cal S}_\mathrm{eff}[x]$ contains a "dissipation" term (the second term) which is nonlocal in imaginary time $\tau$.

\section{Thermal Activation Process}

The thermal activation rate well above the crossover temperature is given by~\cite{rf:Weiss} 
\begin{eqnarray}
\Gamma_\mathrm{TA}
& =&
\frac{\omega_R}{2 \pi}
c_\mathrm{qm}
\exp \left(  - \frac{V_0}{k_B T} \right)
 \\
c_\mathrm{qm}
& =&
 \prod _{n=1}^\infty 
 \frac{\omega_n^2 + \omega_p^2 + \omega_n \hat{\gamma} (\omega_n) }
 {\omega_n^2 - \omega_p^2 + \omega_n \hat{\gamma} (\omega_n) }
 ,
 \end{eqnarray}
where the effective trial frequency $\omega_\mathrm{R}$ is a solution of $\omega_\mathrm{R}^2 + \omega_\mathrm{R}  \hat{\gamma} (\omega_\mathrm{R} )  =\omega_p^2  $, $\omega_n$ is the Matsubara frequency, $\omega_p=\sqrt{2 \pi I_C / \Phi_0 C_J}(1- \gamma^2)^{1/4}$ is the Josephson plasma frequency, 
and $\hat{\gamma}(\omega_n)= (C_S/C_J) |\omega_n| \omega_{LC}^2/(\omega_n^2 +\omega_{LC}^2)$ is the damping function.
Note that the potential barrier height $V_0=(2/27) M \omega_p^2 x_1^2$ ($x_1 = 3 \sqrt{1-\gamma^2}$) is not modified even in the presence of the LC circuit.
Therefore the coupling to the LC circuit only modifies the prefactor of $\Gamma_\mathrm{TA}$.
The effective trial frequency $\omega_\mathrm{R}$ and the quantum correction $c_\mathrm{qm}$ in the prefactor can be calculated analytically, $i.e$.,
\begin{eqnarray}
\omega_\mathrm{R}
&\approx&
\omega_p \sqrt{1-\frac{L_J}{L_S}}
\\
c_\mathrm{qm}^\mathrm{2D}
&\approx&
   \frac{\sinh \left(  \frac{\hbar \beta \omega_p}{2} \right)}
          {\sin \left(  \frac{\hbar \beta \omega_p}{2 } \right)}
          =c_\mathrm{qm}^\mathrm{1D}
\end{eqnarray}
for the non non-adiabatic cases ($\omega_\mathrm{LC} \ll \omega_p$) which is applicable to the MQT experiment in YBCO grain-boundary JJs~\cite{rf:Bauch1,rf:Bauch2}.
In the case of $L_S/L_J \gg 1$, the prefactor of $\Gamma_\mathrm{TA}$ coincides with the result without dissipative effects~\cite{rf:Weiss}.
Therefore we can conclude that the influence of the coupling to the LC circuit on the thermal activation process is quite weak.
So the system behaves as simple 1D systems well above the crossover temperature.
This result is qualitatively consistent with the experimental observation~\cite{rf:Bauch1}.

\section{Macroscopic Quantum Tunneling Process}

The MQT escape rate at zero temperature is defined by $\Gamma_\mathrm{MQT}= \lim_{\beta \to \infty} (2/ \beta) \mathrm{Im} \ln {\cal Z}$~\cite{rf:Weiss}.
In order to determine $\Gamma_\mathrm{MQT}$, we use the bounce techniques.
Then we get 
$
\Gamma_\mathrm{MQT}^\mathrm{2D}
\approx
(\omega_p/2 \pi) \sqrt{120 \pi B } \exp \left(  -B \right)
,
$
where $B = ({\cal S}_0 [x_B]+  {\cal S}_\mathrm{diss} [x_B])/\hbar$ is the bounce exponent, that is the value of the action ${\cal S}_\mathrm{eff}$ evaluated along the bounce trajectory $x_B(\tau)=x_1 \mathrm{sech}(\omega_p \tau/2)$.
In the non-adiabatic case ($\omega_{LC} \ll \omega_p$), we get 
\begin{eqnarray}
\Gamma_\mathrm{MQT}^\mathrm{2D} 
&=&
\frac{\omega_p}{2 \pi} \sqrt{864 \pi \frac{V_0}{\hbar \omega_p}  \left( 1+\frac{\delta M}{M} \right) }
 \exp \left[ 
 - \frac{36}{5} \frac{V_0}{\hbar \omega_p}  \left( 1+\frac{\delta M}{M} \right) 
 \right]
 ,
 \end{eqnarray}
where the "dissipation" correction term is given by $\delta M/M=5L_J/2 L_S$.
In contrast to $\Gamma_\mathrm{TA}$, the coupling to the LC circuit gives rise to an increase of the bounce exponent.
Therefore the two-dimensional nature has large influence on the MQT escape process.

In order to check the validity of our model, we try to compare our result with the experimental data of the switching current distribution at the low temperature regime~\cite{rf:Bauch1}.
The switching current distribution $P(\eta)$ is related to the MQT rate $\Gamma(\gamma, \eta)$ as
\begin{eqnarray}
P(\gamma)=\frac{1}{v}
\Gamma(\gamma) \exp
\left[
- \frac{1}{v}
\int_0^{\gamma} \Gamma(\gamma') d \gamma'
\right]
,
\end{eqnarray}
where $v \equiv \left| d \eta / d t \right| $ is the sweep rate of the external bias current.
From the experiment~\cite{rf:Bauch1}, it is found that the measured value of the full width at half maximum (HMFW) $\sigma$ of $P(\eta)$ is 11.9nA at the temperature-independent MQT regime.
On the other hand, in order to numerically calculate $\sigma$, we need the information of $I_c$, $C_J$, $L_S$, and $C_S$.
The values of $I_c=1.4 \mu$A, $L_s=1.7$nH and $C_S=1.6$pF were directly determined from the TA and the microwave-assisted MQT experiments~\cite{rf:Bauch1,rf:Bauch2}.
Therefore the only fitting parameter is $C_J$.
From the numerical estimation of $\sigma$, we found that $C_J =0.2$ pF gives good agreement with the experimental value of $\sigma$.
The obtained value of $C_J$ is consistent with the estimated value  $C_J \sim 0.1 C_S \sim  0.16$pF based on the geometry of the junction~\cite{rf:Rotoli}.
Therefore, we can conclude that the extended circuit model can quantitatively explain the MQT experiment in the YBCO grain boundary junctions.

\section{Summary}

We have theoretically studied the TA and MQT escape for the Josephson junction coupled with a LC circuit by taking into account the two-dimensional nature of the phase dynamics.
Then we found that the coupling to the LC circuit gives negligible reduction for TA escape rate.
On the other hand, we also found that the MQT escape rate is largely reduced due to the coupling with the LC circuit.
These results are consistent with experimental  data of YBCO Josephson junctions.

\ack
  We would like to thank J. Ankerhold,  M. Fogelstr\"om, G. Johansson, J. R. Kirtley, J. P. Pekola, G. Rotoli, V. Shumeiko, and F. Tafuri  for useful discussions. 
  This work was supported by JST-CREST and the JSPS-RSAS Scientist Exchange Program.\\


\begin{thebibliography}{99}
%
%
%
%
\bibitem{rf:Bauch1}
Bauch T {\it et al} 2005
{\it Phys. Rev. Lett.} {\bf 94} 087003
%
%
\bibitem{rf:Inomata}
Inomata K {\it et al} 2005
{\it Phys. Rev. Lett.} {\bf 95} 107005
%
%
\bibitem{rf:Matsumoto}
Matsumoto T {\it et al} 2007
{\it Supercond. Sci. Technol.} {\bf 20} S10
%
%
\bibitem{rf:Li}
Li S X {\it et al} 2007
{\it Phys. Rev. Lett.} {\bf 99} 037002
%
%
\bibitem{rf:Kashiwaya}
Kashiwaya H {\it et al} 2008
submitted to {\it J. Phys. Soc. Jpn.}
%
%
\bibitem{rf:Bauch2}
Bauch T {\it et al} 2006
{\it Science} {\bf 311} 57
%
%
\bibitem{rf:Jin}
Jin  X Y {\it et al} 2006
{\it Phys. Rev. Lett.} {\bf 96} 177003
%
%
\bibitem{rf:Lombardi}
Lombardi F {\it et al}
2007
{\it IEEE Trans. Appl. Supercond.} {\bf 17} 653
%
%
\bibitem{rf:Rotoli}
Rotoli G {\it et al} 2007
{\it Phys. Rev. B} {\bf 75} 144501
%
%
\bibitem{rf:Weiss}
Weiss U 2008
{\it Quantum Dissipative Systems} (World Scientific, Singapore)
%
%
\bibitem{rf:Kawabata1}
Kawabata S {\it et al} 2004
{\it Phys. Rev. B} {\bf 70} 132505
%
%
\bibitem{rf:Kawabata2}
Kawabata S {\it et al} 2005
{\it Phys. Rev. B} {\bf 72} 052506
%
%
\bibitem{rf:Kawabata3}
Kawabata S {\it et al} 2006
{\it Phys. Rev. B} {\bf 74} 180502(R)
%
%
\bibitem{rf:Kawabata4}
Kawabata S {\it et al} 2007
{\it Phys. Rev. B} {\bf 76} 064505
%
%
\bibitem{rf:Yokoyama}
Yokoyama T {\it et al} 2007
{\it Phys. Rev. B} {\bf 76} 134501
%
%
\bibitem{rf:Kawabata5}
Kawabata S {\it et al} 2007
{\it Supercond. Sci. Technol.} {\bf 20} S6
%
%
\end{thebibliography}
\end{document}